# Instability of the shock wave/ sonic surface interaction

A. Kuzmin

alexander.kuzmin@pobox.spbu.ru

Dept. of Fluid Dynamics

St. Petersburg State University

St. Petersburg

Russia

## Abstract

The work addresses 2D and 3D turbulent transonic flows past a wall with an expansion corner. A curved shock wave is formed upstream of a cylinder located above the corner. Numerical solutions of the Reynolds-averaged Navier-Stokes equations are obtained on fine meshes with a finite-volume solver of the second order accuracy. The solutions demonstrate the existence of adverse free-stream Mach numbers which admit abrupt changes of the shock position at small perturbations. This is explained by an instability of the closely spaced sonic surface and shock wave on the wall.

## Nomenclature

| | |
|---|---|
| $a_\infty$ | − free-stream sound speed |
| $h$ | − height of the plate |
| $H$ | − height of the computational domain |
| $M_\infty$ | − free-stream Mach number |
| $M_{mid}$ | − midvalue of the oscillating free-stream Mach number |
| $p_\infty$ | − static pressure in the free stream |
| $Re$ | − Reynolds number |
| $t$ | − time |
| $T$ | − period of free-stream Mach number oscillations |
| $T_\infty$ | − static temperature in the free stream |
| $U_\infty, V_\infty, W_\infty$ | − free-stream velocity components |
| $x, y, z$ | − non-dimensional Cartesian coordinates |
| $x_c$ | − $x$-coordinate of the expansion corner |
| $x_{out}, y_{out}$ | − coordinates of the lower edge of the outlet |
| $x_{sh}$ | − $x$-coordinate of the shock at $y=0.17$ |
| $y^+$ | − non-dimensional thickness of the first mesh layer on the wall |
| $\alpha$ | − angle of attack |
| $\theta$ | − expansion corner angle |

## 1. Introduction

In the 1990s and 2000s, transonic flow simulations revealed an instability of double supersonic regions on airfoils or flattened bumps in a channel [1-6]. The instability results from an interaction between the shock, which terminates the aft supersonic region, and the sonic line, which is a front of the rear supersonic region. The origin of the instability is seen from considerations of a double supersonic region in the steady inviscid flow. Indeed, the distance $d$ between the sonic line and normal shock on the airfoil (Fig. 1a) decreases as the free-stream Mach number $M_\infty > 1$ gradually increases. However it cannot vanish because the flow is strictly subsonic behind the shock, therefore the shock and sonic line cannot have a common point on the airfoil. As a consequence, when $M_\infty$ exceeds a certain value, the shock jumps downstream and creates a coalescence of the aft and rear supersonic regions, see Fig. 1b. In the 3D flow over wings, the supersonic regions may coalesce either gradually or abruptly, depending on the wing sweep angle [7].

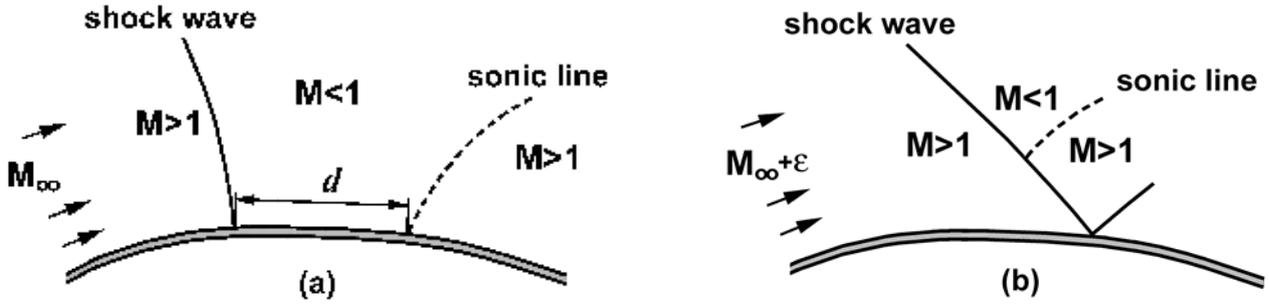

Figure 1. A scheme of an abrupt change of the shock position on a wall under a small increase in the Mach number $M_\infty$.

Recently Kuzmin [8] studied transonic flow in a channel where a shock is formed due to a bend of the upper wall, while the sonic line arises due to an expansion corner of the lower wall. A dependence of the shock wave instability on the velocity profile given at the inlet was discussed. In practice, such a problem occurs, e.g., in supersonic intakes which encounter variations of the incoming flow because of the atmospheric turbulence or a maneuvering flight of aircraft.

In this paper we address a similar problem in which the upper wall is replaced by a cylinder whose axis is normal to the plane $(x,y)$.

## 2. Formulation of the problem and a numerical method

A wall with an expansion corner of 10° is given by the expressions

$$y=0 \text{ at } 0< x \leq x_c \ ; \quad y= -(x-x_c)\tan(10°) \text{ at } x_c < x < x_{out}.$$

Above the wall, there is a circle of radius $r$ whose center resides at a height $h=0.3$ m and has an abscissa of 0.21 m. The circle in the 2D flow formulation is virtually a section of the 3D cylinder with the axis normal to the plane $(x,y)$. In what follows, the Cartesian coordinates $(x,y,z)$ and radius $r$ are non-dimensionalized by $h$, therefore the coordinates of the circle center are $x=0.7$ and $y=1$, see Fig. 2.

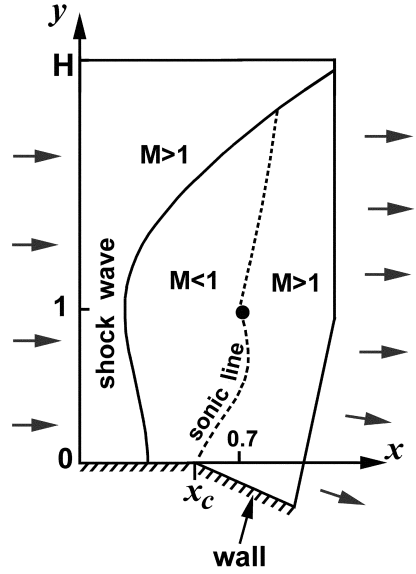

Figure 2.  Sketch of the computational domain and the sonic line location.

The left boundary of the computational domain is set at $x=0$, $0<y<H$. The upper boundary is remote at a distance $H=6$ from the wall in order to eliminate its influence on the flow region between the wall and cylinder. The outlet boundary is constituted by two segments with endpoints

$$x=1.5,\ y=1\ ;\quad x=1.5,\ y=H,$$

and

$$x=x_{out},\ y=y_{out}\ ;\quad x=1.5,\ y=1,$$

where $x_{out}=1$, $y_{out}=-(x_{out}-x_c)\tan(10°)$.

At the outlet, a condition of the supersonic flow regime is imposed. On the left boundary $x=0$, we prescribe the flow velocity, static pressure $p_\infty=100{,}000$ N/m$^2$, and static temperature $T_\infty=250$ K which determines the sound speed $a_\infty=317.02$ m/s. We use the no-slip condition and vanishing heat flux on the wall and cylinder. Initial data are either parameters of the uniform free stream or a flow field calculated for a different Mach number of the incoming flow. The air is treated as a perfect gas whose specific heat at constant pressure is 1004.4 J/(kg K) and the ratio of specific heats is 1.4. We adopt the value of 28.96 kg/kmol for the molar mass, and use the Sutherland formula for the molecular dynamic viscosity.

The free-stream Mach numbers under consideration are $1.117 \leq M_\infty \leq 1.28$, therefore the Reynolds number based on $M_\infty$ and the height $h=0.3$ m is about $8.7 \times 10^6$.

Solutions of the unsteady Reynolds-averaged Navier-Stokes equations were obtained with an ANSYS-15 CFX finite-volume solver of the second order accuracy. An implicit backward Euler scheme was employed for the time-accurate computations. We used a Shear Stress Transport $k$-$\omega$ turbulence model which is known to reasonably predict aerodynamic flows with boundary layer separations [9].

Numerical simulations of 2D flow were performed on hybrid meshes constituted by quadrangles in 39 layers on the wall and cylinder, and by triangles in the remaining region. The non-dimensional thickness $y^+$ of the first mesh layer on the wall and cylinder was less than 1. Apart from the boundary layer region, mesh nodes were clustered in vicinities of the expansion corner and shock wave. Test computations on uniformly refined meshes of approximately $10^5$, $2 \times 10^5$, and $4 \times 10^5$ cells showed that a discrepancy between shock wave coordinates obtained on the second and third meshes did not exceed 1%. Global time steps of $10^{-6}$ s and $2 \times 10^{-6}$ s yielded undistinguishable solutions. That is why we employed meshes of $2 \times 10^5$ cells and the time step of $2 \times 10^{-6}$ s for the study of 2D transonic flow at various free-stream velocities. The root-mean-square CFL number (over mesh cells) was about 3.

Simulations of 3D flow were performed in a domain created by an extrusion of the 2D domain in the $z$-direction from $z=0$ up to $z=1$. A hybrid mesh was constituted by $3.2 \times 10^6$ prisms in 39 layers on the wall, cylinder and side boundaries, and by $18.1 \times 10^6$ tetrahedrons in the remaining region.

The solver was verified by computation of a few commonly used test cases, such as transonic flow over RAE 2822 airfoil [10], ONERA M6 wing [11], and in a channel with a circular-arc bump and a curved shock on the bump [8]. The calculated flow fields were in good agreement with numerical and experimental data available in the literature.

## 3. Shock wave position versus $M_\infty$ for the cylinder of radius $r = 0.01$

First, we suppose the free stream is uniform and parallel to the *x*-axis. Then the *x*- and *y*-components of the inflow velocity are

$$U_\infty = M_\infty a_\infty, \qquad V_\infty = 0 \qquad \text{at } x=0,\ 0<y<H. \tag{1}$$

Numerical simulations of the 2D turbulent flow demonstrated a convergence of the mean parameters to steady states in less than 0.2 s of physical time.

For $M_\infty=1.15$ and the expansion corner located at $x_c=0.6$, the obtained steady flow field exhibits a curved shock, behind which the velocity is subsonic except for a small vicinity of the corner, see Fig. 3a. The vicinity resides below the line $y=0.17$, and there are intersections of the line with the V-shaped contour $M(x,y)=1$. Hereafter the *x*-coordinate $x_{sh}$ of the left intersection will be used to trace the streamwise position of the shock.

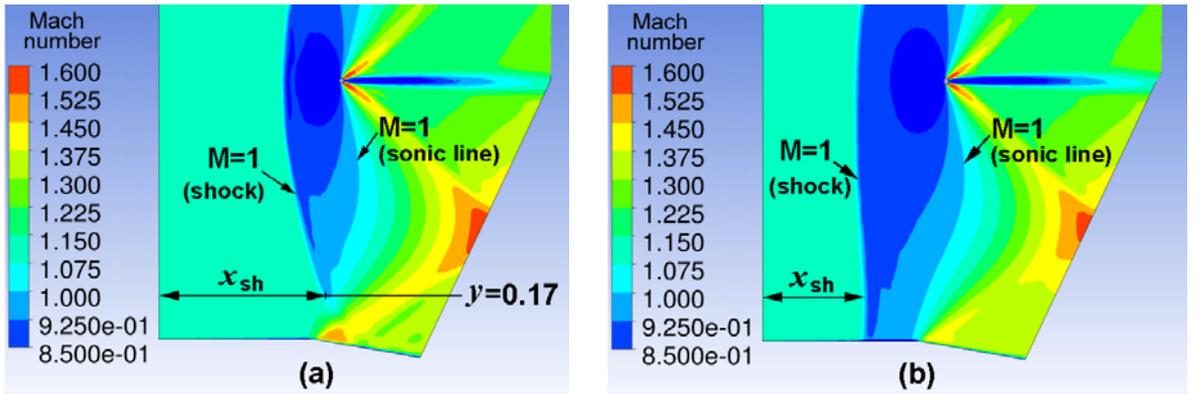

Figure 3. Mach number contours in the flow over the wall with the corner located at $x_c=0.6$: (a) $M_\infty=1.15$, (b) $M_\infty=1.129$.

Computations showed that the shock coordinate $x_{sh}$ gradually decreases as $M_\infty$ decreases from 1.15 to 1.131 (see the upper branch of Plot 3 in Fig. 4). If $M_\infty$ is further decreased to 1.13, then the supersonic region splits, and the relaxation results in a steady state with two supersonic regions separated by a subsonic zone, see Fig. 3b. Therefore the shock coordinate $x_{sh}$ jumps from 0.555 to 0.403. Further decrease of $M_\infty$ to 1.104 entails a shift of the shock wave upstream, a decrease of $x_{sh}$, and an increase of the distance between the shock and the expansion corner. Inversely, if $M_\infty$ increases from 1.104 to 1.13, then the shock gradually shifts downstream, and $x_{sh}$ rises from 0.059 to 0.403. A subsequent increase of $M_\infty$ to 1.131 causes a jump of the shock to the position $x_{sh}=0.555$.

For the expansion corner coordinates $x_c=0.4$ and $x_c=0.5$, computations demonstrated a similar behavior of the shock, see Plots 1 and 2 in Fig. 4, respectively.

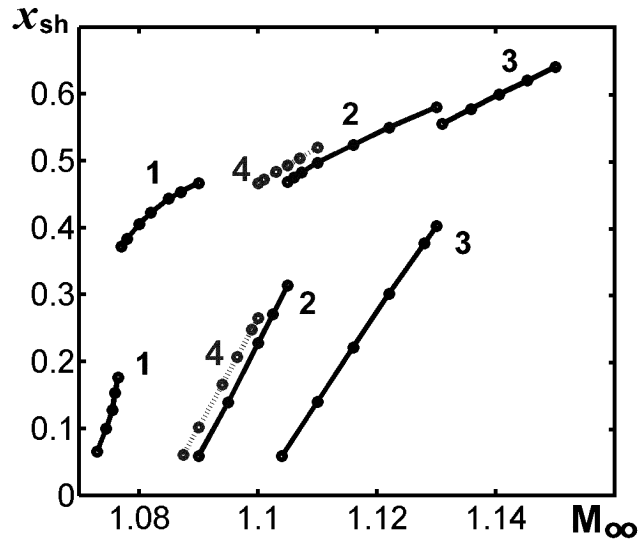

Figure 4. Shock wave coordinate $x_{sh}$ at the height $y=0.17$ versus the free-stream Mach number $M_\infty$ in 2D flow at $r=0.01$ and various locations of the expansion corner: 1 – $x_c=0.4$, 2 – $x_c=0.5$, 3 – $x_c=0.6$ (turbulent flow); 4 – $x_c=0.5$ (inviscid flow).

Now suppose the free-stream Mach number oscillates, while the flow direction remains parallel to the $x$-axis:

$$U_\infty = M_\infty(t)\, a_\infty, \qquad V_\infty = 0 \quad \text{at} \quad x=0,\ 0<y<H,$$

where $M_\infty(t) = (1+\delta \sin(2\pi t/T))\, M_{mid}$.

If $M_{mid}=1.11$ and $\delta=0.0045248$, then $M_\infty(t)$ oscillates between 1.105 and 1.115. For the period $T=1/7$ s and $x_c=0.5$, the numerical simulation showed oscillations of the shock position in the short interval

$$0.470 \le x_{sh}(t) \le 0.528.$$

If $M_{mid}$ is decreased to 1.107, then $M_\infty(t)$ oscillates between 1.102 and 1.112 with the same amplitude. Meanwhile the calculated amplitude of the shock position oscillations is increased by the factor of 2.3:

$$0.368 \le x_{sh}(t) \le 0.501. \tag{2}$$

This is explained by the shock instability and the switching between flow patterns which correspond to the upper and lower parts of Plot 2 in Fig. 4. We notice that interval (2) is yet shorter than the interval $0.261 \le x_{sh} \le 0.506$ determined by Plot 2 for stationary Mach numbers $1.102 \le M_\infty \le 1.112$, because the time $T=0.2$ s is insufficient for accomplishing the flow relaxation to steady states.

## 4. Shock wave position versus $M_\infty$ for the cylinder of radius $r=0.02$

If the radius of cylinder is doubled, then plots $x_{sh}(M_\infty)$ exhibit noticeable hystereses in addition to the discontinuities, and there exist non-unique flow regimes in narrow bands of $M_\infty$, see Fig. 5.

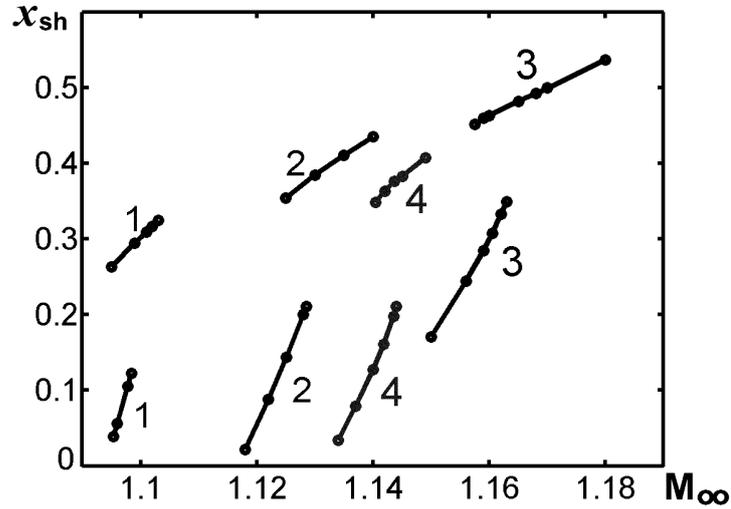

Figure 5. Shock wave coordinate $x_{sh}$ versus $M_\infty$ at $r=0.02$ and various locations of the expansion corner: $1 - x_c=0.3$, $2 - x_c=0.4$, $3 - x_c=0.5$ (2D flow); $4 - x_c=0.4$ (3D flow).

In 3D flow simulations, the side boundaries $z=0$ and $z=1$ are treated as solid walls of a channel. The vanishing $z$-component of the flow velocity $W_\infty = 0$ is added to conditions (1) on the left boundary. For the expansion corner location $x_c=0.4$, the obtained shock position versus $M_\infty$ is illustrated by Plot 4 in Fig. 5. The coordinate $x_{sh}$ of 3D shock is calculated at $y=0.17$ in the midspan section $z=0.5$ of the channel. A comparison of Plots 2 and 4 shows that, though the side walls influence the shock considerably, a jump of the 3D shock is similar to the one in 2D flow. Figure 6 illustrates the shock and sonic surface locations at $M_\infty=1.143$.

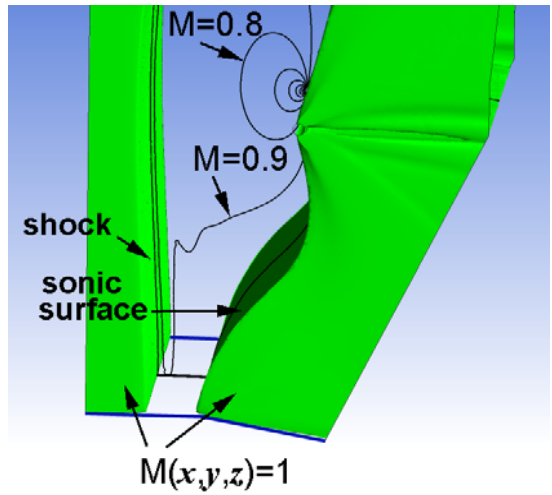

Fig. 6. Surfaces M(*x*,*y*,*z*)=1 in the 3D flow and Mach number contours in the midsection *z*=0.5 at $M_\infty$=1.143, $x_c$=0.4.

When the angle θ=10° of the expansion corner is replaced by θ=13° or 16°, both the shock and sonic surface positions remain the same. The replacement influences flow parameters only downstream of the corner in the domain *y*≤0.7*x*, *x*>0 . There is no boundary layer separation from the wall at 1.09 ≤$M_\infty$≤1.18.

## 5. Conclusion

The numerical simulations of shock wave and sonic line/surface locations near the expansion corner have revealed jumps of the shock position at adverse free-stream Mach numbers. The jumps become stronger when the corner shifts upstream of the cylinder which generates the shock. The phenomenon is true for both turbulent and inviscid flows. 3D flow simulations confirm the findings.

## Acknowledgements

This research was performed using computational resources provided by the Computational Center of St. Petersburg State University (http://cc.spbu.ru). The work was supported by the Russian Foundation for Basic Research under a grant no. 13-08-00288.

# References

1. Jameson A. Airfoils admitting non-unique solutions of the Euler equations, *AIAA Paper*, no. 91-1625, pp. 1-13, 1991.
2. Hafez M. and Guo W. Nonuniqueness of transonic flows, *Acta Mechanica*, Vol. 138, pp. 177-184, 1999.
3. Kuzmin A. Interaction of a shock wave with the sonic line, *Proceedings of IUTAM Symposium Transsonicum IV*, Ed.: H. Sobieczky, Kluwer, pp. 13-18, 2003.
4. Kuzmin A.G. and Ivanova A.V. The structural instability of transonic flow associated with amalgamation/splitting of supersonic regions, *Journal of Theoretical and Computational Fluid Dynamics*, Vol. 18, no. 5, pp. 335-344, 2004.
5. Kuzmin A. Aerodynamic surfaces admitting jumps of the lift coefficient in transonic flight, In "*Computational Fluid Dynamics Review 2010*", World Scientific Publishing, Chapter 22, pp. 543-562, 2010.
6. Kuzmin A. Non-unique transonic flows over airfoils, *Computers and Fluids*, Vol. 63, pp. 1-8, 2012.
7. Kuzmin A. Sensitivity analysis of transonic flow over J-78 wings, *International Journal of Aerospace Engineering*, Hindawi, Vol. 2015, article ID 579343, pp. 1-7, 2015.
8. Kuzmin A. Shock wave instability in a channel with an expansion corner, *International Journal of Applied Mechanics*, Vol. 7, no. 2, 2015.
9. Menter F.R. Review of the Shear-Stress Transport turbulence model experience from an industrial perspective, *International Journal of Computational Fluid Dynamics*, Vol. 23, pp. 305-316, 2009.
10. Kuzmin A. Lift sensitivity analysis for a Withcomb airfoil with aileron deflections, *Progress in Computational Fluid Dynamics*, Vol.15, no.1, pp. 10-15, 2015.
11. Kuzmin A. On the lambda-shock formation on ONERA M6 wing, *International Journal of Applied Engineering Research*, Vol. 9, no. 20, pp. 7029-7038, 2014.